\documentclass[11pt]{amsart}
\usepackage{amsthm}

\usepackage[all]{xy}
\usepackage{amssymb}
\usepackage{enumerate}
\usepackage{mathrsfs}
\usepackage{graphicx}
\usepackage{caption}
\usepackage{subcaption}
\usepackage{float}
\usepackage{epigraph}

\numberwithin{equation}{section}

\newtheorem{thm}{Theorem}[section]

\newtheorem{lem}[thm]{Lemma}

\newtheorem{rem}{Remark}[section]

\newcommand{\eq}[1]{(\ref{#1})}

\renewcommand{\Re}{\operatorname{\rm Re}}
\renewcommand{\Im}{\operatorname{\rm Im}}

\newcommand{\beqast}{\begin{eqnarray*}}
\newcommand{\eqast}{\end{eqnarray*}}
\newcommand{\beqa}{\begin{eqnarray}}
\newcommand{\eqa}{\end{eqnarray}}

\newcommand{\bbe}{\begin{equation}}
\newcommand{\ee}{\end{equation}}

\renewcommand{\Re}{\operatorname{\rm Re}}
\renewcommand{\Im}{\operatorname{\rm Im}}

\newcommand{\bC}{{\mathbb C}}
\newcommand{\bE}{{\mathbb E}}

\newcommand{\bP}{{\mathbb P}}

\newcommand{\bR}{{\mathbb R}}

\newcommand{\bZ}{{\mathbb Z}}

\newcommand{\cF}{{\mathcal F}}

\newcommand{\cG}{{\mathcal G}}

\newcommand{\cL}{{\mathcal L}}

\newcommand{\cC}{{\mathcal C}}

\newcommand{\barX}{{\bar X}}
\newcommand{\uX}{{\underline X}}

\newcommand{\phipq}{{\phi^+_q}}
\newcommand{\phimq}{{\phi^-_q}}

\newcommand{\Om}{{\Omega}}

\newcommand{\al}{\alpha}

\newcommand{\de}{\delta}
\newcommand{\eps}{\epsilon}

\newcommand{\la}{\lambda}

\newcommand{\mum}{\mu_-}
\newcommand{\mup}{\mu_+}

\newcommand{\num}{\nu_-}
\newcommand{\nup}{\nu_+}

\newcommand{\sg}{\sigma}

\newcommand{\om}{\omega}
\newcommand{\omm}{\om_-}
\newcommand{\omp}{\om_+}

\newcommand{\ze}{\zeta}

\newcommand{\ga}{\gamma}
\newcommand{\gap}{\gamma_+}
\newcommand{\gam}{\gamma_-}

\newcommand{\dd}{\partial}

\begin{document}

\title[Simulation of a L\'evy process, its extremum, and hitting time of the extremum]
{Simulation of a L\'evy process, its extremum, and hitting time of the extremum
via characteristic functions}
\author[
Svetlana Boyarchenko and
Sergei Levendorski\u{i}]
{
Svetlana Boyarchenko and
Sergei Levendorski\u{i}}

\begin{abstract}
 We suggest a general framework for simulation of the triplet $(X_T,\barX_ T,\tau_T)$ (L\'evy process, its extremum, and hitting time of the extremum), and, separately, $X_T,\barX_ T$ and pairs $(X_T,\barX_ T)$, $(\barX_ T,\tau_T)$,
 $(\barX_ T-X_T,\tau_T)$, via characteristic functions and conditional characteristic functions.
 The  conformal deformations technique allows one to evaluate  probability distributions, joint probability distributions and conditional probability distributions accurately and fast. 
 For simulations in the far tails of the distribution, we precalculate and store the values of the (conditional) characteristic 
 functions on multi-grids on appropriate surfaces in $\bC^n$, and use these values to calculate the quantiles in the tails.
 For simulation in the central part of a distribution, we precalculate the values of the cumulative distribution
 at points of a non-uniform (multi-)grid, and use interpolation to calculate quantiles.

\end{abstract}

\thanks{
\emph{S.B.:} Department of Economics, The
University of Texas at Austin, 2225 Speedway Stop C3100, Austin,
TX 78712--0301, {\tt sboyarch@utexas.edu} \\
\emph{S.L.:}
Calico Science Consulting. Austin, TX.
 Email address: {\tt
levendorskii@gmail.com}}

\maketitle

\noindent
{\sc Key words:} L\'evy process, extrema of a L\'evy process,  barrier options, Wiener-Hopf factorization, Fourier transform, Laplace transform, 
 Gaver-Wynn Rho algorithm, sinh-acceleration, SINH-regular processes, Stieltjes-L\'evy processes

\noindent
{\sc MSC2020 codes:} 60-08,42A38,42B10,44A10,65R10,65G51,91G20,91G60

\tableofcontents

\section{Introduction}\label{s:intro}
Let $X$ be a one-dimensional L\'evy process on the filtered probability space $(\Om, \cF, \{\cF_t\}_{t\ge 0}, \bP)$
satisfying the usual conditions, and let $\bE$ be the expectation operator under $\bP$. 
Let  $\barX_  t=\sup_{0\le s\le t}X_s$ and $\uX_t=\inf_{0\le s\le t}X_s$ be the supremum
and infimum processes (defined path-wise, a.s.); $X_0=\barX_   0=\uX_0=0$. Let $\tau_T$ be the first time at which $X$ attains its supremum. The joint probability distribution $V(a_1,a_2;T,t):=\bP[X_T\le a_1,\barX_T\le a_2, \tau_T\le t]$, where $a_1\le a_2$, $a_2>0$ and $0<t\le T$,
of the triplet $\chi=(X_T, \barX_ T, \tau_T)$ 
is an important object in insurance mathematics,
structural credit risk models, mathematical finance, buffer size in queuing theory and the prediction of the ultimate supremum and its time in optimal stopping. As it stated in \cite{MijatovicGeomConvSimul21}, for a general L\'evy process,  analytical calculations are extremely challenging, which lead to the development of numerous approximate methods, mostly, Monte Carlo and multi-level Monte Carlo. The literature on the Monte-Carlo simulations is huge but the simulation of probability distributions of L\'evy processes, 
the joint probability distribution of the L\'evy process and its extremum, and more involved probability distributions, remains very difficult. The Monte Carlo simulation of stable L\'evy distributions is extensively studied in the literature
for half a century
 (see, e.g., \cite{ChambersMallowsStuckBW76,SamorodnitskyTaqqu94,Weron96} and the bibliographies therein);
in particular, in the op.cit., one can find exact representations of stable random variables as functions
of normal and exponential variables. For L\'evy processes with the finite second moment, approximation
of small jumps by an additional Brownian motion (BM) component is suggested in \cite{AssRos}, and the BM and compound Poisson components are simulated independently. 
Simulation of the extremum $\barX_ T$ of a L\'evy process
and joint distributions of $(X_T,\barX_T)$ and $(X_T,\barX_ T,\tau_T)$ is more involved, nevertheless,
during the last dozen of years, important advances have been made. See \cite{KuzKyprMC,CGMYsim,
MijatovicDrawdownDuration22,MijatovicGeomConvSimul21,MijatovicStableSupremum23} and the bibliographies therein. As in \cite{AssRos}, subtle probability tools are used to construct an approximation of the process amenable to efficient simulations,
and then the convergence of the simulation algorithm is studied. See, e.g., \cite{MijatovicGeomConvSimul21}, 
where a geometrically convergent algorithm is constructed.  

However, in a number of concrete situations, the theoretical probability arguments are insufficient to estimate  errors of 
algorithms accurately. A theoretically exact representation may fail to simulate the distribution in
tails, as the example on p.46 in \cite{SamorodnitskyTaqqu94}  for stable L\'evy processes on $\bR$ demonstrates. 
For L\'evy processes with exponentially decaying tails of the L\'evy density, the approximation
approach \cite{AssRos} may produce sizable errors if applied to price options with barrier features,
and simulation of the supremum process faces the same difficulties in a somewhat different form. The main source of
difficulties is the qualitatively different behavior of prices of barrier options
in pure jump L\'evy models, probability distributions of $\barX_T$ in particular.  
See \cite{KudrLev09} for numerical examples of errors stemming from the application of the method \cite{AssRos}
to pricing barrier options,  \cite{NG-MBS,early-exercise,BIL,asymp-sens} for asymptotic formulas for prices of barrier options as the underlying approaches the barrier, and \cite{EfficientLevyExtremum,
EfficientStableLevyExtremum,EfficientDoubleBarrier,AltFX,Joint-3} for numerical examples that illustrate the asymptotic results.
Therefore, it is important to develop methods which allow for an efficient error control.  Even in cases when accurate methods are  computationally more expensive than the methods available
in the literature, the former can be used to determine the range of applications where the latter are sufficiently accurate. 

In the paper, we suggest a general methodology to construct methods that are fairly fast and allow for
an efficient error control.
The underlying idea is standard and used in applications to simulation of L\'evy processes on $\bR$ in a number
of publications. If $Z$ is a 1D random variable with continuous cpdf, one can simulate $Z$ sampling a uniformly distributed random variable $U$ on $(0,1)$ and calculating the quantile $F^{-1}(u)$, where $F$ denotes the cumulative distribution function (cpdf) of $Z$, and $u\in (0,1)$ the sample. 
Typically,  explicit formulas for cpdfs are not available but in many cases,
explicit formulas for the characteristic functions are. One can apply the inverse Fourier transform
and evaluate cpdf at points of an appropriate grid and use  interpolation to calculate the quantile. This approach fails in the case of distribution with slowly decaying tails if standard tools such as
the Fast Fourier transform (FFT) or fast Hilbert transform are used as in 
 \cite{glass-liu-07,GlassermanLiu07,glass-liu-10,ChenFengLin,CGMYsim,feng-lin11}. 
In Section  \ref{s:XT}, we explain fundamental difficulties which FFT- and fast HT-based
simulation schemes face, and recall the conformal acceleration method, which allows one to evaluate probability
distribution function (pdf) and cpdf of a L\'evy process very accurately and fast. Then we outline simulation schemes of L\'evy processes
on $\bR$ developed in \cite{SINHregular,ConfAccelerationStable}. The schemes are based on the  precalculation
of  values of $F$ in the central part of the distribution and values of the characteristic function
at points of a grid on a conformally deformed line of integration that are used to evaluate quantiles in the tails.

In the following sections, we explain how the schemes in Section \ref{s:XT} are modified for more difficult situations.
The simulation of the supremum process in Section \ref{s:XTandbarXT} is quite similar. Important new features are: 1) instead
of the characteristic function $e^{-T\psi(\xi)}$ defined on a domain in $\bC$, the Wiener-Hopf factor $\phipq(\xi)$
defined on a domain in $\bC^2$ appears, and we have to use the double inverse Laplace-Fourier transform;
2) efficient methods \cite{Contrarian} for evaluation of the Wiener-Hopf factors and two-dimensional inversion are needed;
3) two-dimensional grids are needed, and two-dimensional arrays stored. The methods developed in \cite{Contrarian} 
allow one to use two-dimensional arrays of  moderate sizes. Simulation of the infimum process is by symmetry,
and simulation of the drawdown $\barX_T-X_T$ is reducible to simulation of $\uX_T$.

The idea of  simulation of joint distributions in Section  \ref{s:sim_joint} is as follows. Suppose, we have two random random variables $X,Y$ with the smooth joint cpdf $F_{X,Y}(x,y)$, and explicit integral representations for  $\dd_1F_{X,Y}(x,y)$ and $F_X(x), p_X(x)$, amenable to fast and accurate calculations. To simulate $(X,Y)$, we take a random sample $(u_1,u_2)$ from 
the uniform distribution $U((0,1)^2)$, and then (1) solve the equation $F_X(x)=u_1$; (2) solve the equation
$\dd_1F_{X,Y}(x,y)= u_2p_X(x)$. For one sample, the sinh-acceleration allows one to find $(x,y)=(x(u_1),y(u_1,u_2))$
fairly fast but for accurate simulations one needs hundreds of thousands if not millions samples.
Therefore, as in Sections \ref{s:XT}-\ref{s:XTandbarXT}, it is necessary to precalculate certain arrays
of values of the probability distributions and characteristic functions. 
In Sections \ref{ss:XT_barXT_tauT} and \ref{ss:drawdown_tauT}, the joint probability distribution
of three random variables appears, and the scheme is modified in the natural fashion.

In Section \ref{s:concl}, we summarize the results of the paper and outline possible extensions.

 \section{Simulation of $X_T$}\label{s:XT}
 \subsection{Classes of processes and general formulas}\label{ss:classes}
 For $\mum<\mup$, $\ga\in (0,\pi)$ and $\ga^-_0<0<\ga^+_0$, define the strip $S_{(\mum,\mup)}:=
 \{\xi\in \bC\ |\ \Im\xi\in (\mum,\mup)\}$ and coni  $\cC_\ga=\{\xi\in \bC\ |\ \mathrm{arg}\,\xi\in (-\ga,\ga)\}$, $
\cC^+_{\ga^-_0,\ga^+_0}:=\{\xi\in \bC\ |\ \mathrm{arg}\, \xi\in (\ga^-_0,\ga^+_0)\}$, 
$\cC^-_{\ga^-_0,\ga^+_0}:=\{\xi\in \bC\ |\ \mathrm{arg}\,\xi\in (\pi-\ga^-_0,\pi-\ga^+_0)\}$, 
$\cC_{\ga^-_0,\ga^+_0}:=\cC^+_{\ga^-_0,\ga^+_0}\cup \cC^-_{\ga^-_0,\ga^+_0}$. 
 In \cite{EfficientAmenable}, we proved that the characteristic exponent $\psi$ of essentially all popular classes of L\'evy processes on $\bR$ bar stable L\'evy processes of index $\al\in (0,2)$ enjoy the following properties
 \begin{enumerate}[(i)]
 \item
 there exists $\mu\in \bR$ s.t.
 \bbe\label{repr_psi}
 \psi(\xi)=-i\mu\xi+\psi^0(\xi),\ \xi\in\bR;
 \ee
\item
there exist $\mum<\mup$ and $\ga^-_0<0<\ga^+_0$ such that 
 $\psi^0$ admits analytic continuation to $S_{(\mum,\mup)}+ (\cC_{\ga^-_0,\ga^+_0}\cup\{0\})$; 
\item
 as $(\cC_{\ga^-_0,\ga^+_0}\ni)\xi\to\infty$, $\Re \psi^0(\xi)> c_\infty |\xi|^\nu$, where $c_\infty>0$.
\end{enumerate}
In \cite{SINHregular}, we gave a more general and detailed definition of a class of SINH-processes enjoying properties
(i)-(iii) with $\mum\le 0\le \mup$, $\mum< \mup$ and $\gam\le 0\le \gap$, $\gam<\gap$.
In \cite{EfficientAmenable}, we defined a class of Stieltjes-L\'evy processes (SL-processes). In order to save space, we do not reproduce the complete set of definitions. Essentially, $X$ is called a (signed) SL-process if $\psi$ is of the form
\bbe\label{eq:sSLrepr}
\psi(\xi)=(a^+_2\xi^2-ia^+_1\xi)ST(\cG^0_+)(-i\xi)+(a^-_2\xi^2+ia^-_1\xi)ST(\cG^0_-)(i\xi)+(\sg^2/2)\xi^2-i\mu\xi, 
\ee
where $ST(\cG)$ is the Stieltjes transform of a (signed) Stieltjes measure $\cG$,  $a^\pm_j\ge 0$, and $\sg^2\ge0$, $\mu\in\bR$.
We call a (signed) SL-process $X$ SL-regular if $X$ is SINH-regular. We proved in \cite{EfficientAmenable} that if $X$ is a (signed) SL-process then $\psi$ admits analytic continuation to the complex plane with two cuts along the imaginary axis, and
if $X$ is a SL-process, then, for any $q>0$, equation $q+\psi(\xi)=0$ has no solution on $\bC\setminus i\bR$.
We also proved that all popular classes of L\'evy processes bar the Merton model and Meixner processes are regular SL-processes, with $\ga_\pm=\pm \pi/2$;
the Merton model and Meixner processes are regular signed SL-processes, and $\ga_\pm=\pm \pi/4$.
 For lists of SINH-processes and SL-processes, with calculations of the order and type,
see \cite{EfficientAmenable}.

In the case of stable L\'evy processes, the strip degenerates into $\bR$, and $\psi^0$ admits analytic continuation
from $(0,+\infty)$ to $\cC^+_{\ga^-_0,\ga^+_0}$ and from $(-\infty,0)$ to $\cC^-_{\ga^-_0,\ga^+_0}$. In the case of asymmetric stable L\'evy processes of index $\al=1$ and similar L\'evy processes with exponentially decaying tails,
either $\mup=0$ or $\mum=0$, and the most efficient type of conformal deformations (sinh-acceleration or exponential acceleration in the case of stable L\'evy processes) are not always applicable. See \cite{ConfAccelerationStable} for details. In the case of Variance Gamma processes (VGP), condition (iii) holds with $\ln|\xi|$ instead of $|\xi|^\nu$,
and longer grids are needed to satisfy a given error tolerance. 

   Let $t>0$ and $a\in\bR$. Assume that (i)-(iii) hold with $\mup>0$. Then the pdf $p_{X}(t;a)$ and cpdf 
   $F_X(t;a)$ of $X$ can be calculated as 
 \beqa\label{pdf1}
 p_X(t,a)&=&\frac{1}{2\pi}\int_{\Im\xi=\omp}e^{i(-a+t\mu)\xi-t\psi^0(\xi)}d\xi,\\\label{cpdf1}
 F_X(t,a)&=&\frac{1}{2\pi}\int_{\Im\xi=\omp}\frac{e^{i(-a+t\mu)\xi-t\psi^0(\xi)}}{-i\xi}d\xi,
 \eqa
 where $\omp\in (\mum,\mup)$ in \eq{pdf1} and $\omp\in (0,\mup)$ in \eq{cpdf1} are arbitrary. In \eq{cpdf1}, we can pass to the limit as $\omp\downarrow 0$, and, using the residue theorem, obtain
 \bbe\label{cpdf1b}
 F_X(t,a)=\frac{1}{2}+\frac{1}{2\pi}\mathrm{v.p.}\int_{\bR}\frac{e^{i(-a+t\mu)\xi-t\psi^0(\xi)}}{-i\xi}d\xi,
 \ee
 where $\mathrm{v.p.}$ denotes the Cauchy principal value. In the case of stable L\'evy processes, we
 can use \eq{pdf1} with $\omp=0$ and \eq{cpdf1b}.

One is tempted to use either the fast Fourier transform (FFT) or
fast Hilbert transform (fast HT), which allow one to calculate the values $F(t,x)$ at
all points of a uniformly spaced grid $x_1<x_2<\cdots<x_M$ faster than point-by-point, especially if the number of points is large.
However, 
for an accurate simulation, the uniform grid
must be very fine, and if the tails decay slowly, then the length of the grid must be very large. In the result, even
grids of dozens of millions of points  can be insufficient. See \cite{SINHregular} for examples.

\subsection{Sinh-acceleration and exponential acceleration}\label{ss:conf_accel}
\begin{figure}
\begin{center}
\includegraphics[width=9.4cm,height=9.5cm]{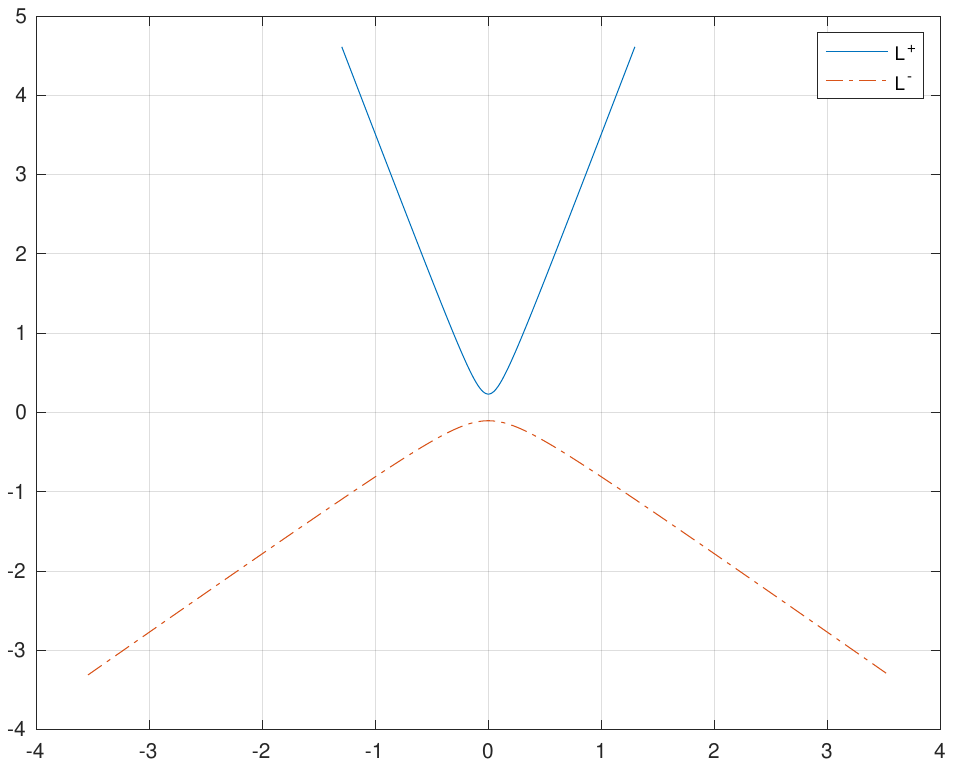}    
\caption{$L^+=\cL^+_{-0.25,0.5,5\pi/12}$, $L^-=\cL^-_{0.25,0.5,-\pi/4}$.}
\label{TwoCurveFrame}
\end{center}
\end{figure}

For $\om_1\in \bR$, $b>0$ and $\om\in (-\pi/2,\pi/2)$, define the map $\chi_{\om_1,b,\om}:\bC\mapsto \bC$ by
$\chi_{\om_1,b,\om}(y)=i\om_1+b\sinh(i\om+y)$ and deform the line of integration
in \eq{pdf1}-\eq{cpdf1} into the curve $\cL_{\om_1,b,\om}=\chi_{\om_1,b,\om}(\bR)$:
\beqa\label{pdf2}
 p_X(t,a)&=&\frac{1}{2\pi}\int_{\cL_{\om_1,b,\om}}e^{i(-a+t\mu)\xi-t\psi^0(\xi)}d\xi,\\\label{cpdf2}
 F_X(t,a)&=&\frac{1}{2\pi}\int_{\cL_{\om_1,b,\om}}\frac{e^{i(-a+t\mu)\xi-t\psi^0(\xi)}}{-i\xi}d\xi.
 \eqa
If $\om>0$ (resp., $\om<0$), the wings of the curve $\cL_{\om_1,b,\om}$ point upwards (resp., downwards). When we wish to indicate that $\om>0$ (resp., $\om<0$), we  write   $\cL^\pm_{\om_1,b,\om}$ or $\cL^\pm$. 
The parameters $\om_1,b,\om$ are chosen so that $\cL_{\om_1,b,\om}\subset S_{(\mum,\mup)}+(\cC_{\ga^-_0,\ga^+_0}\cup\{0\})$ and $0\not\in \cL_{\om_1,b,\om}$. In particular, $\om\in (\ga^-_0,\ga^+_0)$. Furthermore, it is advantageous (and in the case $\nu<1$ necessary) to choose $\om$ so that if $-a+t\mu\neq 0$ the oscillating factor $e^{i(-a+t\mu)\xi}$ decays as $\xi\to\infty$ along $\cL_{\om_1,b,\om}$. Hence, if $-a+t\mu>0$ (resp. $-a+t\mu<0$), we choose $\om\in (0,\ga^+_0)$ (resp., $\om\in (\ga^-_0, 0)$),
and the choice $\om=\ga^+_0/2$ (resp., $\om=\ga^-_0/2$) is approximately optimal. If $-a+t\mu=0$, the choice
$\om=(\ga^+_0+\ga^-_0)/2$ is approximately optimal. 
On the RHS of \eq{cpdf1}, the curve of integration must remain above 0 in the process of deformation, hence, 
$\om_1+b\sin\om>0$. If $-a+t\mu<0$ and $\om<0$, it is advantageous to push the curve down and cross the pole so that the contour of integration
 is in the lower half-plane:
\bbe\label{cpdf1low}
F_X(t,a)=1+\frac{1}{2\pi}\int_{\cL_{\om'_1,b,\om}}\frac{e^{i(-a+t\mu)\xi-t\psi^0(\xi)}}{-i\xi}d\xi,
\ee
where $\om'_1+b\sin\om<0$. In this paper, we use the curves $\cL^+$ and $\cL^-$ lying in the upper and lower half-planes, respectively.
See 
Fig.~\ref{TwoCurveFrame}. 
However, in some cases, it may be useful to allow for the curve of the type $\cL^+$ (resp., $\cL^-$) to be below
(resp., above) the origin. See \cite{Contrarian} for examples.

The deformation being made, we change the variable $\xi=i\om_1+b\sinh(i\om+y)$ in \eq{pdf2}-\eq{cpdf2}
\beqa\label{pdf3}
 p_X(t,a)&=&\frac{b}{2\pi}\int_{\bR}e^{i(-a+t\mu)\xi(y)-t\psi^0(\xi(y))}\cosh(i\om+y)dy,\\\label{cpdf3}
 F_X(t,a)&=&\frac{b}{2\pi}\int_{\bR}\frac{e^{i(-a+t\mu)\xi(y)-t\psi^0(\xi(y))}}{-i\xi(y)}\cosh(i\om+y)dy,
 \eqa
 where $\xi(y)=\chi_{\om_1,b,\om}(y)$,
apply the infinite trapezoid rule, and truncate the sum
\beqa\label{pdf4}
 p_X(t,a)&\approx&\frac{\ze b}{2\pi}\sum_{|j|\le N}e^{i(-a+t\mu)\xi(j\ze)-t\psi^0(\xi(j\ze))}\cosh(i\om+j\ze),\\\label{cpdf4}
 F_X(t,a)&\approx&\frac{\ze b}{2\pi}\sum_{|j|\le N}\frac{e^{i(-a+t\mu)\xi(j\ze)-t\psi^0(\xi(j\ze))}}{-i\xi(j\ze)}
 \cosh(i\om+j\ze).
 \eqa
 Using the symmetry, the number of terms can be decreased almost two-fold.
The integrands on the RHSs of  \eq{pdf3} and \eq{cpdf3} being analytic in a strip $S_{(-d,d)}$, where $d>0$ depends on $\psi^0$ and the parameters of the deformation, the discretization error of the infinite trapezoid rule decays as $\exp[-2\pi d/\ze]$.
In more detail, let $f$ be analytic in the strip
$S_{(-d,d)}$ and decay at infinity sufficiently fast so that
\[\lim_{A\to \pm\infty}\int_{-d}^d |f(i a+A)|da=0,\]
and $H(f,d):=\|f\|_{H^1(S_{(-d,d)})}$ defined by
\[
H(f,d)=\lim_{a\downarrow -d}\int_\bR|f(i a+ y)|dy+\lim_{a\uparrow d}\int_\bR|f(i a+y)|dy
\]
is finite. The following key lemma is proved in \cite{stenger-book} using the heavy machinery of sinc-functions. For a simple proof, see
\cite{paraHeston}.
\begin{lem}[\cite{stenger-book}, Thm.3.2.1]
The error of the infinite trapezoid rule 
\bbe\label{inftrap}
\int_\bR f(y)dy\approx \ze\sum_{j\in \bZ} f(j\ze),
\ee
where $\ze>0$,
admits an upper bound 
\bbe\label{Err_inf_trap}
{\rm Err}_{\rm disc}\le H(f,d)\frac{\exp[-2\pi d/\ze]}{1-\exp[-2\pi d/\ze]}.
\ee
\end{lem}
Once
an approximately bound for $H(f,d)$ is derived, it becomes possible to choose $\ze$ to satisfy the desired error tolerance. 
Typically, $H(f,d)$ increases with $d$, and the upper bound for $d$ which can be
achieved with an appropriate choice of the deformation is either $-\ga^-_0/2$ or $\ga^+_0/2$.  In many simple cases, one can solve the problem of the minimization of the RHS of \eq{Err_inf_trap} but it is more efficient to choose $\ze$
using 
\eq{Err_inf_trap}  with $d$ which is 0.8-0.9 of the upper bound for the half-width of the strip of analyticity.
The error of the truncation of the infinite sum can be estimated via integrals as well, and
approximate recommendations for a choice of the parameters of the numerical scheme are easy to derive. 
The new integrand decays as a double exponential function,
hence, the complexity of the numerical scheme is of the order of $E\ln E$, where $E=\ln (1/\eps)$ and $\eps>0$ is the error tolerance.  

In the case of  stable L\'evy processes, we separate the integral into two (over $\bR_\pm$). Using the  symmetry, one may evaluate the integral over $\bR_+$ only. We rotate the ray of integration, make the exponential change of the variable $\xi=e^{i\om+y}$, and apply the simplified trapezoid rule. Since the integrand does not decay
fast as $y\to-\infty$,  fast calculations are possible only if the integrand admits a convenient asymptotic expansion as $\xi\to 0$ remaining in a cone in the right half-plane. We use appropriate asymptotic expansions as $y\to-\infty$ to increase the
rate of the convergence of the simplified trapezoid rule. The complexity of the scheme is of the order of $E^{1+a}$, where $a>0$ depends on the properties of the asymptotic expansion used. See \cite{ConfAccelerationStable,ConfAsymp,Sinh} for details in applications to the evaluation of special functions and stable distributions. 

\begin{rem}\label{rem:alt} {\rm
Alternatively, we approximate the characteristic exponent of a stable L\'evy process by a function analytic in the union of
a strip and cone. For a realization of this idea and error bound for approximations, see \cite{NewFamilies}.
For instance, $|\xi|^\al$ is approximated by $(\la^2+\xi^2)^{\al/2}$, and the sinh-acceleration allows one
to evaluate the integral with the accuracy of the order E-12 using $\la=10^{-8}$ at a small CPU cost.
The alternative approach to the numerical evaluation of stable L\'evy distributions is almost as efficient as
the exponential change of variables unless the Blumenthal-Getoor index $\al$ is rather close to 0.
At the same time, it becomes unnecessary to formulate the simulation procedures  for the pdf and cpdf of extrema of stable L\'evy processes and joint distributions in terms of exact formulas derived in \cite{EfficientStableLevyExtremum}
for stable L\'evy processes.}
\end{rem}

\subsection{Simulation scheme}\label{ss:simul_XT}
Fix $t>0$ and temporarily denote $F(x)=F(t,x)$. In \cite{SINHregular,ConfAccelerationStable}, for wide classes of L\'evy processes
on $\bR$, we developed efficient schemes for the evaluation of quantiles $F^{-1}(u)$. 
We outline the scheme for the case of L\'evy processes with exponentially decaying tails of the L\'evy density;
the method in \cite{ConfAccelerationStable} for stable L\'evy processes is a modification.
\begin{enumerate}[I]
\item 
Instead of FFT or fast HT, we use the conformal deformations method (typically, the most efficient
sinh-acceleration method can be applied) which allows one to evaluate pdfs and cpdfs with
precision E-13 in milliseconds (using double precision arithmetic, MATLAB and Mac of moderate characteristics) for wide regions in the parameter space for a chosen family of processes;  calculations in the tails are especially efficient. 
\item
Using the sinh-acceleration method, we precalculate
the values of $F(x)$ at the points of a  grid $x_1<x_2<\cdots<x_M$ such that $F(x_1)<\de$ and $F(x_M)>1-\de$ for a chosen small $\de>0$, e.g., $\de=0.01$ or $\de=0.005$. If a sample $u$ from the uniform distribution is in the range
$[\de, 1-\de]$, we find $k$ such that $F(x_k)\le u< F(x_{k+1})$, and use the precalculated values $F(x_k)$ 
and $F(x_{k+1})$ and linear interpolation, as in \cite{glass-liu-07}, to fund $F^{-1}(u)$.
We can also precalculate $p(x_k), k=1,2,\ldots, M$, and use interpolation procedures of higher order, for instance,
solving the equation $F(x_k)+(x-x_k)p(x_k)=u$ or $F(x_{k+1})+(x-x_{k+1})p(x_{k+1})=u$
(the former equation is preferred  if $F(x_{k+1})-u<u-F(x_k)$). Approximations of higher order are very efficient if $x_k$ is far from the peak. 
The advantage of the sinh-acceleration method as compared to the FFT or fast HT based methods is that we can use non-uniform grids; and for small $T$, many distributions have very high peaks. If the peak is high, and the slopes are steep, it is advantageous to use a fine grid in the vicinity of the peak and sparse grid farther from the peak. 
\item
The third improvement of the standard scheme is the evaluation of $F$ in the tails. When FFT or fast HT are used, then if $u<\de$ (resp., $u>1-\de$) one sets $F^{-1}(u)=x_1$ (resp., $F^{-1}(u)=x_M$). This simplification produces large errors if the tails exhibit very slow decay and a multi-step Monte-Carlo procedure is used: relatively small errors at each time step accumulate.
We precalculate and store  the values of $\xi$, $\cosh(\xi)$ and $e^{-T\psi^0(\xi)}$ at points of the grid on
 a conformally deformed
line of integration in the standard Fourier inversion formula ({\em flat iFT}) which are needed to accurately evaluate $F(x)$ and $p(x)$ for $x$ in a neighborhood of $-\infty$, and use these values to solve the equation $F(x)=u$ if $u<\de$, with the initial approximation $x=x_1$ ($p=F'$ is needed if Newton's method is applied). Similarly, we store values of $\xi$, $\cosh(\xi)$ and $e^{-T\psi^0(\xi)}$ at points of another grid needed to solve the equation
$F(x)=u$ if $u>1-\de$, with the initial approximation $x=x_M$. Naturally, this additional step requires additional CPU time each time $u\not\in[\de,1-\de]$,
but if $\de$ is chosen sufficiently small, then the probability that additional time needs to be spend is small,
and, furthermore, the size of arrays that needs to be stored and time needed to calculate the quantile decreases with 
$\de$. 
\item
The next trick allows us to decrease the number of points smaller still. Instead of the equation
$F(x)=u$, we solve the equation $f(x)=v$, where $f(x)=\ln F(x)$ and $v=\ln u$. Since $f$ is more regular than $F$,
the same approximations work better. 
\end{enumerate}

\section{Simulation of $\barX_T$ and $\barX_T-X_T$}\label{s:XTandbarXT}
\subsection{General formulas}
Let $q>0$, and let $T_q$ be an exponentially distributed random variable of mean $1/q$, independent of $X$.
For $T>0$, the Laplace transforms of $F_{\barX_T}(h)=\bP[\barX_T\le h]$ and $F_{\uX_T}(h)=\bP[\uX_T\ge h]$
can be expressed in terms of 
the Wiener-Hopf factors $\phipq(\xi)=\bE[e^{i\xi \barX_{T_q}}]$ and $\phimq(\xi)=\bE[e^{i\xi \uX_{T_q}}]$.
If $\psi(\xi)$ admits analytic continuation to a strip $S_{(\mum,\mup)}$, then $\phi^\pm_q(\xi)$ enjoy the following 
properties (for the proof, see \cite{NG-MBS,paired,EfficientLevyExtremum}).
\begin{enumerate}[(a)]
\item
 For any $\sg_0>0$, 
there exist $<\mum<\sg_-<0<\sg_+<\mup$ such that 
$q+\psi(\xi)\not\in (-\infty,0]$ on $\{(q,\xi)\in \bC^2\ |\ \Re q>\sg_0, \sg_-<\Im\xi<\sg_+\}$;
\item
$\phipq(\xi)$ (resp., $\phimq(\xi)$) admits analytic continuation to
$\{(q,\xi)\in \bC^2\ |\ \Re q>\sg_0, \Im\xi>\sg_-\}$ (resp., $\{(q,\xi)\in \bC^2\ |\ \Re q>\sg_0, \Im\xi>\sg_-\}$)
\item
Let $\Re q>\sg_0, \Im\xi>\sg_-$. Then, for any $\omm\in (\mum, \Im \xi)$,
\bbe\label{phip1}
\phipq(\xi)=\exp\left[\frac{1}{2\pi i}\int_{\Im\eta=\omm}\frac{\xi\ln(q/(q+\psi(\eta))}{\eta(\eta-\xi)}d\eta\right].
\ee
\item
Let $\Re q>\sg_0, \Im\xi<\sg_+$. Then, for any $\omp\in (\Im\xi, \mup)$,
\bbe\label{phim1}
\phimq(\xi)=\exp\left[-\frac{1}{2\pi i}\int_{\Im\eta=\omp}\frac{\xi\ln(q/(q+\psi(\eta))}{\eta(\eta-\xi)}d\eta\right].
\ee
\end{enumerate}
Using the standard Fourier/Laplace technique and the properties (b)-(d), one easily obtains
(see, e.g., \cite{NG-MBS,paired}): 
for $h>0$ and $t>0$
\beqa\label{pdfbarXt}
p_{\barX}(t,h)&=&\frac{1}{2\pi i}\int_{\Re q=\sg}dq\,\frac{e^{qt}}{q}\frac{1}{2\pi}\int_{\Im\xi=\omm}
e^{-ih\xi}\phipq(\xi)d\xi,\\\label{cpdfbarXt}
F_{\barX}(t,h)&=&1+\frac{1}{2\pi i}\int_{\Re q=\sg}dq\,\frac{e^{qt}}{q}\frac{1}{2\pi}\int_{\Im\xi=\omm}
e^{-ih\xi}\frac{\phipq(\xi)}{-i\xi}d\xi,
\eqa
and for $h<0$ and $t>0$
\beqa\label{pdfuXt}
p_{\uX}(t,h)&=&\frac{1}{2\pi i}\int_{\Re q=\sg}dq\,\frac{e^{qt}}{q}\frac{1}{2\pi}\int_{\Im\xi=\omp}
e^{-ih\xi}\phimq(\xi)d\xi,\\\label{cpdfuXt}
F_{\uX}(t,h)&=&\frac{1}{2\pi i}\int_{\Re q=\sg}dq\,\frac{e^{qt}}{q}\frac{1}{2\pi}\int_{\Im\xi=\omp}
e^{-ih\xi}\frac{\phimq(\xi)}{-i\xi}d\xi.
\eqa
The Wiener-Hopf factors decaying slow at infinity, the numerical evaluation of 
the integrals on the RHSs of \eq{phip1}-\eq{phim1} and \eq{pdfbarXt}-\eq{cpdfuXt} using FFT or fast HT is very inefficient. If conditions (i)-(iii) in Section \ref{s:XT} hold, and $q>0$, then the integrals w.r.t.  $\xi$ and $\eta$ can be efficiently evaluated using
the sinh-acceleration. The exterior integrals on the RHSs of \eq{pdfbarXt}-\eq{cpdfuXt}
can be evaluated using the Gaver-Stehfest (GS) algorithm or more efficient Gaver-Wynn-Rho (GWR) algorthm.
Both use the values of the integrand for positive $q$'s only; 
for the sinh-acceleration to be applicable to the Bromwich integral,
$\psi$ must satisfy additional conditions.

\subsection{Evaluation of the Wiener-Hopf factors  for  $q>0$}\label{extr_sinh_WHF}
 In \cite{Contrarian} (see also \cite{EfficientLevyExtremum,Joint-3}), we proved the following lemma
\begin{lem}\label{lem:WHF-SINH}
Let conditions (i)-(iii) of Section \ref{ss:classes} hold. Then $\exists$ $\sg>0$ s.t. $\forall$ $q>\sg$, 
\begin{enumerate}[(i)]
\item
$\phipq(\xi)$ admits analytic continuation to $i(\mum,+\infty)+i(\cC_{\pi/2-\ga^-_0}\cup\{0\})$. For any $\xi\in i(\mum,+\infty)+i(\cC_{\pi/2-\ga^-_0}\cup\{0\})$, and any contour 
$\cL^-_{\om_1  ,b,\om}\subset i(\mum,\mup)+(\cC_{\ga^-_0,\ga^+_0}\cup\{0\})$ lying below $\xi$,
\bbe\label{phipq_def}
\phipq(\xi)=\exp\left[\frac{1}{2\pi i}\int_{\cL^-_{\om_1  ,b,\om}}\frac{\xi \ln (q/(q+\psi(\eta))}{\eta(\eta-\xi)}d\eta\right];
\ee
\item
$\phimq(\xi)$ admits analytic continuation to $i(-\infty,\mup)-i(\cC_{\pi/2+\ga^+_0}\cup\{0\})$. For any $\xi\in i(\-\infty,\mup)-i(\cC_{\pi/2+\ga^+_0}\cup\{0\})$, and any contour 
$\cL^+_{\om_1  ,\om,b}\subset i(\mum,\mup)+(\cC_{\ga^-_0,\ga^+_0}\cup\{0\})$ lying above $\xi$,
\bbe\label{phimq_def}
\phimq(\xi)=\exp\left[-\frac{1}{2\pi i}\int_{\cL^+_{\om_1  ,\om,b}}\frac{\xi \ln (q/(q+\psi(\eta))}{\eta(\eta-\xi)}d\eta\right].
\ee
\end{enumerate}
\end{lem}

The integrals are efficiently evaluated making the change of variables $\eta=\chi_{\om_1,b,\om}(y)$ and applying the simplified trapezoid rule.
In the process of deformation, the expression $1+\psi(\eta)/q$ may not assume value zero. In order to avoid complications stemming from analytic continuation to an appropriate Riemann surface, it is advisable to ensure that $1+\psi(\eta)/q\not\in(-\infty,0]$.
 Thus, if $q>0$  and $X$ is an SL-process, 
any $\om\in (0,\pi/2)$ is admissible in \eq{phipq_def}, and any $\om\in (-\pi/2,0)$ is admissible in \eq{phimq_def}. 
Recall that  only positive $q$'s are used in the GS or GWR algorithms.
\subsection{Decomposition of the Wiener-Hopf factors}\label{WHF_decomp}
In the remaining part of the paper, we assume that the Wiener-Hopf factors $\phi^\pm_q(\xi), q>0,$ admit
the representations $\phi^\pm_q(\xi)=a^\pm_q+\phi^{\pm\pm}_q(\xi)$
where $a^\pm_q\ge 0$, and $\phi^{\pm\pm}_q(\xi)$ satisfy the following bounds
\beqa\label{WHFdecayP}
 |\phi^{++}_q(\xi)|&\le & C_+(q)(1+|\xi|)^{-\nup},\ \Im \xi\ge \mum,\\
 \label{WHFdecayM}
 |\phi^{--}_q(\xi)|&\le & C_+(q)(1+|\xi|)^{-\num},\ \Im \xi\le \mup,
 \eqa
 where $\nu_\pm>0$ and $C_\pm(q)>0$ are independent of $\xi$. These conditions are satisfied for all popular classes of L\'evy processes
 bar the driftless Variance Gamma model.

The following more detailed properties of the Wiener-Hopf factors are established in \cite{NG-MBS,BLSIAM02,barrier-RLPE}
for the class of RLPE (Regular L\'evy processes of exponential type); the proof for SINH-regular processes is the same only $\xi$ is allowed to tend to $\infty$ not only in the strip of analyticity but in the union of a strip and cone. See \cite{BIL,asymp-sens,paired} for the proof of the statements below for
several classes of SINH-regular processes (the definition of the SINH-regular processes formalizing properties used in \cite{BIL,asymp-sens,paired} was suggested in \cite{SINHregular} later). The contours  in Lemma \ref{lem:atoms} below
are in a domain of analyticity s.t. $q-i\mu\xi\neq 0$ and $1+\psi^0(\xi)/(q-i\mu\xi)\not\in (-\infty,0]$. These restrictions on the contours are needed when $\psi^0(\xi)=O(|\xi|^\nu)$ as $\xi\to\infty$ in the domain of analyticity and $\nu<1$. Clearly, in this case, for sufficiently large $q>0$, the condition holds.
In the case of RLPE's, the contours of integration in the lemma below are straight lines in the strip of analyticity. 

\begin{lem}\label{lem:atoms}
Let $\mum<0<\mup$,  $q>0$, and conditions (i)-(iii) of Section \ref{ss:classes} hold.  Then
\begin{enumerate}[(1)]
\item
if  $\nu\in [1,2]$ or $\nu\in (0,1)$ and the drift  $\mu=0$,  neither
 $\barX_{T_q}$ nor $\uX_{T_q}$ has an atom at 0, and $\phi^\pm_q(\xi)$ admit bounds \eq{WHFdecayP} and 
 \eq{WHFdecayM}, 
 where $\nu_\pm>0$ and $C_\pm(q)>0$ are independent of $\xi$;
 \item
  if  $\nu\in [0+,1)$  and $\mu>0$, then
 \begin{enumerate}[(a)]
 \item
 $\barX_{T_q}$ has no atom at 0 and $\uX_{T_q}$ has an atom  $a^-_q\de_0$ at zero, where
 \bbe\label{eq:cmqp}
a^-_q=\exp\left[-\frac{1}{2\pi i}\int_{\cL^+_{\om_1  ,b,\om}}\frac{\ln((1+\psi^0(\eta)/(q-i\mu\eta))}{\eta}d\eta\right],
\ee
and $\cL^+_{\om_1  ,b,\om}$ is a contour as in Lemma \ref{lem:WHF-SINH} (ii), lying above $0$;
 \item
for $\xi$ and $\cL^-_{\om_1  ,b,\om}$ in Lemma \ref{lem:WHF-SINH} (i), $\phipq(\xi)$ admits the representation
 \bbe\label{phip1finvar}
 \phipq(\xi)=\frac{q}{q-i\mu\xi}\exp\left[\frac{1}{2\pi i}\int_{\cL^-_{\om_1  ,b,\om}}\frac{\xi\ln(1+\psi^0(\eta)/(q-i\mu\eta))}{\eta(\xi-\eta)}d\eta\right],
 \ee
and  satisfies the bound \eq{WHFdecayP} with $\nup=1$;
 \item
 $\phimq(\xi)=a^-_q+\phi^{--}_q(\xi)$, where $\phi^{--}_q(\xi)$ satisfies 
 \eq{WHFdecayM} with arbitrary $\num\in (0,1-\nu)$.
 \end{enumerate}
  \item
  if  $\nu\in [0+,1)$  and $\mu<0$, then
 \begin{enumerate}[(a)]
 \item
 $\uX_{T_q}$ has no atom at 0 and $\barX_{T_q}$ has an atom $a^+_q\de_0$ at zero, where
 \bbe\label{eq:cpqp}
a^+_q=\exp\left[\frac{1}{2\pi i}\int_{\cL^-_{\om_1  ,b,\om}}\frac{\ln((1+\psi^0(\eta)/(q-i\mu\eta))}{\eta}d\eta\right],
\ee
and $\cL^-_{\om_1  ,b,\om}$ is a contour as in Lemma \ref{lem:WHF-SINH} (i), lying below $0$;
 \item
for $\xi$ and $\cL^+_{\om_1  ,b,\om}$ in Lemma \ref{lem:WHF-SINH} (ii), $\phimq(\xi)$ admits the representation
 \bbe\label{phim1finvar}
 \phimq(\xi)=\frac{q}{q-i\mu\xi}\exp\left[-\frac{1}{2\pi i}\int_{\cL^+_{\om_1  ,b,\om}}\frac{\xi\ln(1+\psi^0(\eta)/(q-i\mu\eta))}{\eta(\xi-\eta)}d\eta\right],
 \ee
and
 satisfies the bound \eq{WHFdecayM} with $\num=1$;
 \item
 $\phipq(\xi)=a^+_q+\phi^{++}_q(\xi)$, where $\phi^{++}_q(\xi)$ satisfies  
 \eq{WHFdecayP} with arbitrary $\nup\in (0,1-\nu)$.
 \end{enumerate}
 \end{enumerate}
\end{lem}

\subsection{Analytic continuation of the Wiener-Hopf factors w.r.t. $q$}\label{ss:WHF-anal}
To apply the sinh-acceleration to the Bromwich integral, we need to allow for analytic continuation of
the Wiener-Hopf factors to domains of the form $\sg+(\cC_{\pi/2+\om_\ell}\cup\{0\})$, where $\sg>0$ and $\om_\ell>0$.
In \cite{Contrarian} (see also \cite{EfficientLevyExtremum,Joint-3}), we showed that  if
 either $\nu\ge 1$ or $\nu\in (0,1)$ and
$\mu=0$, this is possible. The lemma below is Lemma 2.9 in \cite{Joint-3}.

\begin{lem}\label{lem:der_WHF_q_SINH}
Let conditions (i)-(iii) of Section \ref{ss:classes} hold, and either the order $\nu\in [1,2]$ or  $\nu\in (0,1)$ and the drift is 0. Then
there exist $(\mu'_-,\mu'_+)\subset(\mum,\mup)$, $\mu'_-<0<\mu'_+$, cone $\cC_{\ga'_-,\ga'_+}\subset \cC_{\ga^-_0,\ga^+_0}$, $\ga'_-<0<\ga'_+$, and $\sg_0>0, \om_L\in (0,\pi/2)$ such that
\begin{enumerate}[(a)]
\item
for all $q\in \sg_0+\cC_{\pi/2+\om_L}$ and $\xi\in i[\mu'_-,\mu'_+]+(\cC_{\ga'_-,\ga'_+}\cup \{0\})$,
\bbe\label{good_bound_psiq}
q+\psi(\xi)\not\in  (-\infty,0];
\ee
\item
$\phipq(\xi)$ admits analytic continuation to $(\sg_0+\cC_{\pi/2+\om_L})\times (i(\mu,+\infty)+i(\cC_{\pi/2-\ga'_-}\cup\{0\}))$
and obeys the bounds
\beqa\label{good_bound_phipq}
|\phipq(\xi)|&\le& C_+(|q|^{1/\nu}+|\xi|)^{-\nup},\\\label{good_bound_phipq_der}
|\dd_q^m\dd^n_\xi\phipq(\xi)|&\le& C_{+,m,n}(|q|^{1/\nu}+|\xi|)^{-\nup}|q|^{-m}(1+|\xi|)^{-n},\ n,m\in \bZ_+,
\eqa
where $C_+, C_{+,m,n}$ are independent of $q,\xi$;
\item
$\phimq(\xi)$ admits analytic continuation to $(\sg_0+\cC_{\pi/2+\om_L})\times (i(-\infty,\mup)-i(\cC_{\pi/2+\ga'_+}\cup\{0\}))$
and obeys the bounds
\beqa\label{good_bound_phimq}
|\phipq(\xi)|&\le &C_-(|q|^{1/\nu}+|\xi|)^{-\num},\\\label{good_bound_phimq_der}
|\dd_q^m\dd^n_\xi\phipq(\xi)|&\le & C_{-,m,n}(|q|^{1/\nu}+|\xi|)^{-\num}|q|^{-m}(1+|\xi|)^{-n},
\eqa
where $C_+, C_{+,m,n}$ are independent of $q,\xi$.
\end{enumerate}
\end{lem}

\subsection{Evaluation of pdf and cpdf of $\barX_T$ 
 using the sinh-acceleration}\label{extr_sinh}
 As we proved in \cite{EfficientLevyExtremum} (see also \cite[Sect. 3]{Joint-3}), 
 we can replace $\phipq(\xi)$ on the RHS of \eq{cpdfbarXt}  with $\phi^{++}_q(\xi)$
and $\phimq(\xi)$ on the RHS of \eq{cpdfuXt} with $\phi^{--}_q(\xi)$. If $\phipq(\xi)\neq \phi^{++}_q(\xi)$,
then the inner integral on the RHS of \eq{cpdfbarXt} does not converge absolutely; the one with
 $\phi^{++}_q(\xi)$ in place of $\phipq(\xi)$ does.  The advantage of the replacement of
 $\phimq(\xi)$ by $\phi^{--}_q(\xi)$ is the same. Under conditions (i)-(iii), we 
 we can deform the inner contour into a contour of the form $\cL^-_{\om_1  ,b,\om}$:
  \bbe\label{cpdfbarXt2}
F_{\barX}(t,x)=1+\frac{1}{2\pi i}\int_{\Re q=\sg}\frac{e^{qt}}{q}\frac{1}{2\pi}\int_{\cL^-_{\om_1  ,b,\om}}e^{-i\xi h}\frac{\phi^{++}_q(\xi)}{-i\xi}d\xi,
\ee
make the corresponding sinh-change of variables,
 and apply the simplified trapezoid rule. For each $q$ used in a numerical method for the evaluation of the Bromwich integral, 
the error tolerance of the order of E-12-E-13 can be satisfied using the simplified trapezoid rule with
 150-300 terms (the number depends on the properties of $\psi$, the opening angle of the sector of analyticity especially). 

\begin{figure}
\begin{center}
\includegraphics[width=9.5cm,height=9.5cm]{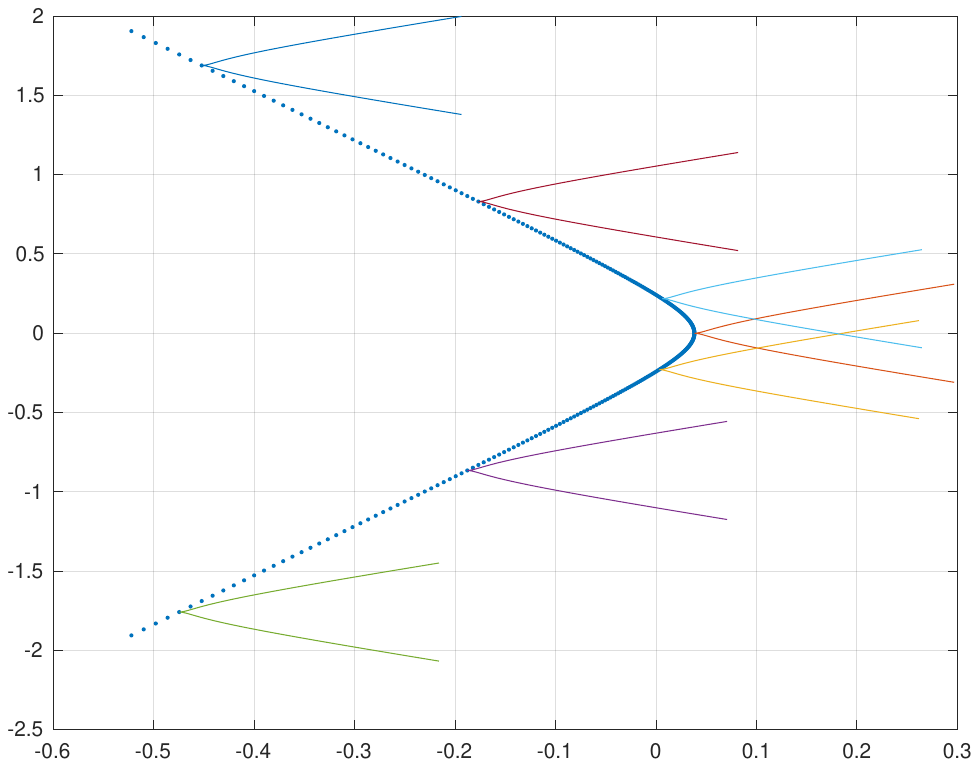}    
\caption{Dotted line: $\cL_{L}$. Solid lines: curves $q+\psi(\cL^-)$ for several values of $q\in \cL_L$}
\label{PsiQm.pdf}
\end{center}
\end{figure}

If either $\nu\ge 1$ or $\nu\in (0,1)$ and
$\mu=0$ then, to calculate the outer integral, we apply the sinh-acceleration or  
 summation by parts in the infinite trapezoid rule (see \cite{Contrarian,Joint-3} for the algorithm) and truncate the sum. The error tolerance of
 the order of E-12 (resp., E-14) can be satisfied using a truncated sum with 150-200 (resp., 200-250) terms. We can also apply the GWR algorithm with $2M=16$ terms but then the best accuracy that can be achieved is of the order of E-07 unless high precision arithmetics and $2M>16$ are used. The GWR algorithm can be used in all cases when (i)-(iii) of Section \ref{ss:classes} hold.
Thus,  if either $\nu\ge 1$ or $\nu\in (0,1)$ and
$\mu=0$, we recommend to apply the sinh-acceleration to the outer integral as well:  
\bbe\label{cpdfbarXt3}
F_{\barX}(t,h)=1+\frac{1}{2\pi i}\int_{\cL^L_{\sg,b_\ell,\om_\ell}}\frac{e^{qt}}{q}\frac{1}{2\pi}\int_{\cL^-_{\om_1  ,b,\om}}e^{-i\xi h}\frac{\phi^{++}_q(\xi)}{-i\xi}d\xi,
\ee Above, $\cL^L_{\sg,b_\ell,\om_\ell}=\chi^L_{\sg,b_\ell,\om_\ell}(\bR)$ and $\chi^L_{\sg,b_\ell,\om_\ell}(y)=\sg+ib_\ell\sinh(i\om_\ell+y)$.
 The parameters are chosen so that, for all $(q,\xi)$ arising in the process of deformations, $q+\psi(\xi)\not\in (-\infty,0]$, and $\xi\neq 0$. 
If $\nu\in (1,2]$ or $\nu\in (0,1]$ and $\mu=0$, the crucial parameters $\gam<\om<0$ and $0<\om_\ell<\pi/2$  must satisfy $\max\{1,\nu\}|\om|<\pi/2-\om_\ell$ (if $\nu=1$ and $\mu\neq 0$, the condition is more involved). 
If  $\max\{1,\nu\}|\om|<\pi/2-\om_\ell$,
it is straightforward to show (see  \cite{Contrarian,EfficientLevyExtremum}) that there exist $\om_1  ,b, \sg, b_\ell$ such that for all $(q,\xi)$ arising in the process of deformations, $q+\psi(\xi)\not\in (-\infty,0]$. See Fig.~\ref{PsiQm.pdf} for illustration.
  
 In \eq{pdfbarXt}, $\phipq(\xi)$ can be replaced with $\phi^{++}_q(\xi)$ as well, and sinh-deformations made:
\bbe\label{pdfbarXt3}
p_{\barX}(t,h)=\frac{1}{2\pi i}\int_{\cL^L_{\sg,b_\ell,\om_\ell}}\frac{e^{qt}}{q}\frac{1}{2\pi}\int_{\cL^-_{\om_1  ,b,\om}}e^{-i\xi h}\phi^{++}_q(\xi)d\xi.
\ee
 Similarly, for the cpdf of the infimum process, we have
 \bbe\label{cpdfuXt3}
F_{\uX}(t,h)=\frac{1}{2\pi i}\int_{L^L_{\sg,b_\ell,\om_\ell}}dq\,\frac{e^{qt}}{q}\frac{1}{2\pi}\int_{\cL^+_{\om_1  ,b,\om}}
e^{-ih\xi}\frac{\phi^{--}_q(\xi)}{-i\xi}d\xi.
\ee

\subsection{Simulation of $\barX_T$}\label{ss:simul_barXT} Temporarily, denote $F(x)=F_{\barX}(T,x)$,
$p(x)=p_{\barX}(T,x)$. The scheme is a straightforward modification of the scheme of simulation of $X_T$.
\begin{enumerate}[I]
\item
choose a small $\de>0$, 
find $F^{-1}(\de)$, choose a non-uniform grid $0=h_1<h_2<\cdots<h_M=F^{-1}(\de)$, dense in a neighborhood of 0 and sparse far from 0, and set $F(x_1)=0$;
\item precalculate
$F(h_k)$ and $p(h_k)$, $k=2,\ldots, M$, using the GWR algorithm and sinh-acceleration if $\nu\in (0,1)$ and
$\mu\neq 0$, and the sinh-acceleration in the Bromwich integral and inverse Fourier transform if either $\nu\in [1,2]$ or
$\nu\in (0,1)$ and $\mu=0$;
\item
if $\nu\in (0,1)$ and
$\mu\neq 0$, then, for each $q$ used in the GWR algorithm, and for each point on the sinh-deformed curve
in $\bC_\xi$-space, precalculate the values of the integrands in the formulas for $F(h)$ and $p(h)$ (bar the factor
$e^{-ih\xi}$) needed for efficient evaluation of $F(h)$ and $p(h)$ in the region $h>h_M$;
\item
if either $\nu\in [1,2]$ or
$\nu\in (0,1)$ and $\mu=0$, then, for each pair $(q,\xi)$ on the Cartesian product of curves in $\bC^2$ used for the double sinh-acceleration,  precalculate the values of the integrands in the formulas for $F(h)$ and $p(h)$ (bar the factors
$e^{-ih\xi}$ and $e^{qT}$) needed for efficient evaluation of $F(h)$ and $p(h)$ in the region $h>h_M$;
\item
take a random sample $u$ from the uniform distribution, and
\begin{enumerate}[(1)]
\item
if $u\le 1-\de$,  find $k$ such that
$F(h_k)<u\le F(h_{x+1})$, and use the arrays precalculated at step II and an interpolation procedure of choice to solve the equation  $F(h)=u$. This step
is essentially the same as in the case of the simulation   of $X_T$;
\item
if $u>1-\de$, use the arrays precalculated at step III or IV (depending on $\nu$ and $\mu$) to solve the equation
$F(h)=u$, with the initial approximation $h=h_M$.
\end{enumerate}
\end{enumerate}

\subsection{Simulation of the drawdown}\label{ss:simul_drawdown}
Let $T>0$ be fixed. For any $q>0$, the random variables $\barX_{T_q}-X_{T_q}$ and $-\uX_{T_q}$ are identical in law,
therefore, for any $a>0$, the Laplace transforms of 
 $\bP[\barX_T-X_T\ge a]$ and $\bP[\uX_T\le -a]$ coincide, and $\bP[\barX_T-X_T\ge a]=\bP[\uX_T\le -a]$,
a.e. For processes that we consider, $\bP[\uX_T\le -a]$ is a continuous function of $a\in (0,+\infty)$,
hence, $\bP[\barX_T-X_T\ge a]=\bP[\uX_T\le -a]$ for all $a>0$. The simulation procedure of $\uX_T$ is the evident mirror reflection of the simulation procedure of $\barX_T$.

\section{Simulation of joint distributions}\label{s:sim_joint}
For the sake of brevity, we consider the case when the sinh-accelartion in the Bromwich integral
can be made. The formulas and schemes can be adjusted to the case when the GWR algorithm is applied
in the same vein as in the case of simulation of the supremum process.

\subsection{Simulation of the pair $(X_T,\barX_T)$}\label{ss:joint_pair1}
In \cite{EfficientLevyExtremum}, we derived the following representation for
the joint probability distribution $F_{X,\barX}(a,h)=\bP[X_T\le a, \barX_T\le h]$,  $h>0$, $a\in (-\infty,h]$, of the L\'evy process and its supremum:
\beqa\label{FXbarX}
F_{X,\barX}(T;a,h)&=&\frac{1}{2\pi}\int_{\cL_{\om_1,b,\om}}\frac{e^{i(-a+T\mu)\xi-t\psi^0(\xi)}}{-i\xi}d\xi\\\nonumber
&+& \frac{1}{(2\pi)^3 i}\int_{L_{\sg,b_\ell,\om_\ell}}dq\,\frac{e^{qT}}{q}
\int_{\cL^-}  d\eta\, e^{-ih\eta}\phi^{++}_q(\eta)
  \int_{\cL^+}d\xi
\, \frac{e^{i\xi(h-a)}\phi^{--}_q(\xi)}{\xi(\xi-\eta)},
 \eqa
 where 
 \begin{enumerate}[(a)]
 \item
 $L_{\sg,b_\ell,\om_\ell}$ is a sinh-deformed contour in the Bromwich integral,
 \item
  $\cL_{\om_1,b,\om}$ is a sinh-deformed contour above $0$ such that. $\om\ge 0$ if $-a+\mu T\ge 0$ and $\om\le 0$ if $-a+\mu T\le 0$;
  \item
 $\cL^+$ (resp., $\cL^-$) is a sinh-deformed contour in the upper (resp., lower) half-plane;
 \item
 $q+\psi(\xi'), q+\psi(\eta), q+\psi(\xi)\not\in (-\infty,0]$ for all $q\in L_{\sg,b_\ell,\om_\ell}$,
 $\xi'\in  \cL_{\om_1,b,\om}$, $\eta\in \cL^-$, $\xi\in \cL^+$, and this property holds in the process of deformation
 of the initial straight lines of integration.
 \end{enumerate} 
 It is easy to prove that, for $a$ fixed, the integral remains absolutely convergent after the differentiation w.r.t. $h$ under
the integral sign. Hence, for $a<h$, we have
\bbe\label{FXbarXder}
\dd_h F_{X,\barX}(T;a,h)=\frac{1}{(2\pi)^3 i}\int_{L_{\sg,b_\ell,\om_\ell}}dq\,\frac{e^{qT}}{q}
\int_{\cL^-}  d\eta\, e^{-ih\eta}\phi^{++}_q(\eta)
  \int_{\cL^+}d\xi
\, \frac{e^{i\xi(h-a)}\phi^{--}_q(\xi)}{-i\xi}.
 \ee
 For $x<h$, the conditional distribution $\bP[X_T<x\ |\ \barX_T=h]$ is given by
 $\bP[X_T<x\ |\ \barX_T=h]=\dd_h F_{X,\barX}(T;x,h)/p_{\barX}(T,h)$, therefore,
 the $u_1$-quantile $x$ of $\bP[X_T<x\ |\ \barX_T=h]$ can be found solving the equation
 \bbe\label{cond_quantile_XbarX}
 \dd_h F_{X,\barX}(T;x,h)=u_1p_{\barX}(T,h).
 \ee
 To simulate the pair $(X_T,\barX_T)$, it is sufficient to simulate the pair $(\barX_T,X_T|\barX_T)$, equivalently,
 for a random sample $(u_1,u_2)$ from the uniform distribution $U((0,1)^2)$, solve the system
 of equations 
 \bbe\label{barXh}
 F_{\barX}(T,h)=u_2
 \ee
 and \eq{cond_quantile_XbarX}.
 Probably, it is optimal to solve \eq{barXh}
 first and then the equation \eq{cond_quantile_XbarX} but it is feasible that one can design faster algorithms
 for the simultaneous solution of the system.
 
 To solve the system, we use \eq{cpdfbarXt3}, \eq{pdfbarXt3} and \eq{FXbarXder}.
 
 \subsubsection{The safest albeit slowest simulation scheme A}\label{sss:  simulXbarX1}
 \begin{enumerate}[I]
 \item
 Represent $(0,1)^2$ as a disjoint union of a finite number of rectangular sets $U_j$, $j=1,2,\ldots, N$ (some are semi-infinite), such that,  for
 each $j$,  one can use the same grids on the curves in the dual spaces to evaluate 
 $F_{\barX}(T,h)$, $p_{\barX}(T,h)$,  and $ \dd_h F_{X,\barX}(T;x,h)$ for $(x,h)\in U_j$
sufficiently accurately. 
 \item
 For each $j$, choose the curves in $q$-, $\xi-$, $\eta-$ and $\xi'$-spaces and grids on the curves
 which can be used to evaluate $F_{\barX}(T,h)$, $p_{\barX}(T,h)$  and $ \dd_h F_{X,\barX}(T;x,h)$ for $(x,h)\in U_j$
with the desired accuracy. 
\item
For each $j$, precalculate  all factors in the formulas \eq{cpdfbarXt3}, \eq{pdfbarXt3} and \eq{FXbarXder}
bar exponential factors at the points of  chosen multi-grids in the dual spaces.
\item
For each $j$, calculate $F_{\barX}(T,h)$, $p_{\barX}(T,h)$  and $ \dd_h F_{X,\barX}(T;x,h)$
at the vertices of $U_j$.
\item
For a random sample $(u_1,u_2)$ from the uniform distribution $U((0,1)^2)$, using the results obtained on Step IV,
find $U_j$ such that $u_2\in pr_2 U_j$ and $u_1\in pr_1 U_j$, where $pr_\ell U$ denotes the projection of $U$ on the $\ell$-th coordinate.
\item
Use the arrays precalculated at Step III to solve the system \eq{barXh},
  \eq{cond_quantile_XbarX}. 
\end{enumerate}
 
 \subsubsection{The simulation scheme B}\label{sss:  simulXbarX2}
 \begin{enumerate}[I]
 \item
 Choose a small $\de>0$,  and a fine grid $(=0)h_1<h_2<\cdots <h_M$ such that $F_{\barX}(T,h_M)\ge 1-\de$.
 Contrary to the simulation scheme for $\barX_T$, the grid must be sufficiently fine so that 
 the interpolation on $[h_2,h_M]$ is possible and the values  $p_{\barX}(T,h)$ and $F_{X_T|\barX_T=h}(x)$ for $h\in (0,h_1)$ can be
 sufficiently accurately approximated by $p_{\barX}(T,h_1)$ and $F_{X|\barX=h_1}(T,x)$.
 \item
 For $m=1,2,\ldots, M$, calculate $p_{\barX}(T,h_m)$ and $F_{\barX}(T,h_m)$.
 \item
 For each $m=2,3,\ldots, M$, find $x_{m,1}$ and $x_{m,M_m}$ such that 
 $F_{X_T|\barX_T=h_m}(x_{m,1})<\de$ and $F_{X_T|\barX_T=h_m}(x_{m,M_m})>1-\de$,
 and construct a fine grid $x_{m,1}<x_{m,2}<\cdots<x_{m,M_m}$.
 \item
 For each pair $(m, \ell), m=1,2,\ldots, \ell=1,2,\ldots, M_m$, choose the curves in the dual space and grids
 on the chosen curves sufficient to evaluate $F_{X|\barX}(T,x,h_m)$ for $x\in (x_{m,\ell}, x_{m,\ell+1})$ with the desired accuracy.
 Precalculate the arrays in the dual space needed to evaluate $F_{X|\barX}(T;x,h_m)$.
 \item
 Precalculate $F_{\barX}(T,h_m)$ and $F_{X|\barX}(T,x_\ell,h_m)$, $m=1,2,\ldots, \ell=1,2,\ldots, M_m$.
 \item
 Separate $(0,1)^2\setminus (\de,1-\de)\times (0,1-\de)$ into a disjoint union of rectangular sets $U_j$ (some are semi-infinite),
 and, for each $U_j$, precalculate the same arrays as in Scheme A.
 \item
 For a random sample $(u_1,u_2)$ from the uniform distribution,
 \begin{enumerate}[(1)]
 \item
 if $(u_1,u_2)\in (0,1)^2\setminus  (\de,1-\de)\times (0,1-\de)$, find $(x(u_1,u_2), h(u_2))$ using Scheme A;
 \item
 if $(u_1,u_2)\in  (\de,1-\de)\times (0,1-\de)$, use the results obtained at Steps II and IV to
 find $m$ and $\ell, \ell'$ such that $F_{\barX}(T,h_m)\le u_2<F_{\barX}(T,h_{m+1})$ and 
$F_{X|\barX}(T,x_\ell,h_m)\le u_1<F_{X|\barX}(T,x_{\ell+1},h_m)$, $F_{X|\barX}(T,x_{\ell'},h_{m+1})\le u_1<F_{X|\barX}(T,x_{\ell'+1},h_{m+1})$. Then
\begin{itemize}
\item
using interpolation and the values $p_{\barX}(T,h_m)$, $F_{\barX}(T,h_m)$ and $p_{\barX}(T,h_{m+1})$, 
$F_{\barX}(T,h_{m+1})$, find an approximate solution $h(u_2)$ of the equation $F_{\barX}(T,h)=u_2$.
\item
using linear interpolation and the values $F_{X|\barX}(T,x_\ell,h_m)$ and 
$F_{X|\barX}(T,x_{\ell+1},h_m)$, find an approximation to the solution of the equation
$F_{X|\barX}(T,x,h_m)=u_2$. Denote the approximation $x(h_m)$;
\item
using linear interpolation and the values $F_{X|\barX}(T,x_{\ell'},h_{m+1})$ and 
$F_{X|\barX}(T,x_{\ell'+1},h_{m+1})$, find an approximation to the solution of the equation
$F_{X|\barX}(T,x,h_{m+1})=u_2$. Denote the approximation $x(h_{m+1})$;
\item
calculate the approximation 
\[
x(u_1,u_2)=x(h_m)+\frac{h(u_2)-h_m}{h_{m+1}-h_m}(x(h_{m+1})-x(h_m)).
\]
\end{itemize}
 \end{enumerate}
 \end{enumerate}

\subsection{Simulation of the pair $(\barX_ T,\tau_T)$}\label{ss:barXT_tauT}
The probability $\bP[\barX_T\le h, \tau_T\le t]$, $t\le T$, is the price of the first touch digital with the upper barrier
$h$ and maturity date $t$. The formula for the latter with the integration over the lines $\{\Re q=\sg\}$ and $\{\Im\xi=\om\}$ was derived in \cite{NG-MBS}; the replacement of $\phipq(\xi)$ with $\phi^{++}_q(\xi)$
and sinh-deformation are justified exactly as in the case of the joint cpdf of $(X_T,\barX_T)$. We have
\[
\bP[ \tau_T\le t, \barX_T\le h]=\frac{1}{2\pi i}\int_{\cL^L_{\sg,b_\ell,\om_\ell}}dq\frac{e^{qt}}{q}\frac{1}{2\pi}
\int_{\cL^-_{\om_1,b,\om}}d\xi\,e^{-i\xi h}\frac{\phi^{++}_q(\xi)}{i\xi},
\]
and the differentiation under the integral sign can be justified to obtain
\[
\dd_h\bP[ \tau_T\le t, \barX_T= h]=-\frac{1}{2\pi i}\int_{\cL^L_{\sg,b_\ell,\om_\ell}}dq\frac{e^{qt}}{q}\frac{1}{2\pi}
\int_{\cL^-_{\om_1,b,\om}}d\xi\,e^{-i\xi h}\phi^{++}_q(\xi).
\]
The conditional probability distribution is
$
\bP[ \tau_T\le t\ |\ \barX_T= h]=\dd_h\bP [\tau_T\le t, \barX_T= h]/p_{\barX_T}(h)$,
and the simulation schemes for $(X_T,\barX_T)$ are modified in the straightforward manner.

\subsection{Simulation of the triplet $(X_T, \barX_T,\tau_T)$}\label{ss:XT_barXT_tauT}
\begin{enumerate}[I]
\item
Take a random sample $(u_1,u_2,u_3)$ from the uniform distribution $U((0,1)^3)$.
\item
Using $u_2$, find $h=h(u_2)$ as the solution of $F_{\barX}(T,h)=u_2$.
\item
Using a simulation procedure in Section \ref{ss:joint_pair1}, calculate $x(T; u_1,u_2)$.
\item
Using the modification of a  simulation procedure in Section \ref{ss:joint_pair1}, calculate
$t(h(u_2),u_3)$.
\end{enumerate}

\subsection{Simulation of the pair $(\barX_T -X_T,\tau_T)$}\label{ss:drawdown_tauT}
Use the procedure in Section \ref{ss:XT_barXT_tauT}.

\section{Conclusion}\label{s:concl}
In the paper, we described  general schemes of simulation of a L\'evy process $X$, its extrema, the drawdown, and several joint distributions. The main elements are efficient procedures for the evaluation of pdfs, cpdfs and
conditional cpdfs of $X_T, \barX_T, \uX_T$, $(X_T,\barX_T)$, $(\barX_T,\tau_T)$, using the sinh-acceleration technique
in the case of processes with exponentially decaying tails of the L\'evy densities and exponential changes of variables in the case
of stable L\'evy processes. The resulting algorithms are more efficient than FFT- and fast HT-based algorithms.
The technique  is applicable if the characteristic exponent admits
analytic continuation to a cone around $\bR$; in the case of processes with exponentially decaying tails,
the characteristic exponent admits analytic continuation to a strip around $\bR$. The technique is used in \cite{SINHregular,ConfAccelerationStable} to evaluate
the probability distributions of L\'evy processes, and in \cite{Contrarian,EfficientLevyExtremum,EfficientStableLevyExtremum} to evaluate the probability distributions of the supremum process and joint probability distributions of $(X_T,\barX_T)$. In the context of the efficient evaluation
of  probability distributions, the contribution of the paper is two-fold. First, we derive explicit formulas for
conditional cpdf amenable to efficient calculations. Secondly, using the idea from \cite{NewFamilies},
we suggest to approximate the characteristic exponent of a stable L\'evy process with an appropriate function which 
is analytic in the union of a strip and cone (e.g., $|\xi|^\al$ is approximated by $(\al^2+\xi^2)^{\al/2}$, where $\la>0$ is very small, e.g., $\la=10^{-8}$), and apply the sinh-acceleration technique. The technique is efficient even for very small
lambdas. 
To calculate the quantiles, we follow the schemes of \cite{SINHregular,ConfAccelerationStable}. In the center of a distribution, we precalculate the values of pdf and cpdf at points of a grid which is fine
near the peak and sparse far from the peak. The sinh-acceleration technique allows us to easily evaluate pdf and cpdf at point of non-uniformly spaced grids. To calculate the quantiles in the tails of distributions, we precalculate
values of the expressions in the integrands bar the exponential factors at points of grids or multi-grids used
in the sinh-acceleration formulas, and use these values to solve the equations for the quantiles. 

The latter trick can be used in the following situation. 
In many cases, one needs to evaluate the expected value of stochastic expressions of the form
$G(X_{t_j}, \barX_{t_j}, \tau_{t_j})$, $t_1<t_2<\cdot<t_n$. An important example are barrier/lookback options
with discrete monitoring and time-dependent barriers. The standard approach is to simulate the process and use simulated trajectories
to approximate the expected value.  Several main blocks of the scheme of the paper can be
used to evaluate such expectations 
faster and more accurately than using the standard multi-step Monte-Carlo simulation procedure.  
To this end, we can separate time moments in several groups such that the expectations in each can be calculated
using the same sinh-deformed curves, grids and precalculated arrays in the state space.
In the algorithms of the present paper, the only change is needed: to evaluate the distribution and conditional distribution of a $t_j$-term,
we precalculate the values of the characteristic exponent $\psi^0(\xi)$ but not the values of the characteristic function
$e^{-T\psi^0(\xi)}$. 


\end{document}